# Quasi-Random Lattice Model for Electrolyte Solutions: State of Art and Future Perspectives.

Elsa Moggia

Applied Electromagnetism Group

DITEN (Department of Naval, Electric and Telecommunication Engineering)

University of Genoa, Via Opera Pia 11A, Genoa, Italy

e-mail elsa.moggia@unige.it; elsa@biosafetyengineering.it;

## INTRODUCTION

In this work, the Quasi-Random Lattice (QRL) model [1-5] is summarized and critically discussed, in order to outline its potentialities and limitations, in perspective of future developments.

QRL primarily focuses on the mean activity coefficient $\gamma_\pm$ (molal scale) of ionic solutions, the model having first been developed in order to provide practical equations to calculate $\gamma_\pm$, that is, equations able to involve a minimal number of unknown or unpredictable quantities. The research was motivated by a criticism emerged from literature, that is, rigorous models, although formally powerful in terms of their thermodynamic consistency or adherence to paradigms of Solution Theory, are counterbalanced by unavoidable use of many arbitrary parameters (often purely hypothetical or experimentally not accessible). Conversely, simplified approaches usually work well when applied to thermodynamic properties for which they have specifically been developed, however their accuracy is often reduced if they are applied to a wider set of thermodynamic functions, due to limitations on their internal consistency caused by imposition of theoretical approximations. Critical discussion and comparison among theories and models for ionic solutions are available in many excellent works [6-8], and will no longer be replicated in the following. In these introductory remarks, we limit ourselves to present some graphical results (Figures from 1 to 4) in order to outline current potentialities of QRL in comparison with other models, while theoretical aspects will be discussed later on. For further comparison with literature, also see Refs. [1-5, 9-16] (DHX [1, 4, 9]; HNC [1, 9, 10]; MC [1, 9, 10]; MSA [1, 10]; MPB [1, 10]; IPBE [1, 4, 11]; RIE [4, 12]; BIMSA [4, 13]; SiS [4, 14]; REUNIQUAC [4, 15]; Bahe-Varela [4, 16]). Pitzer Theory [17, 18] or other approaches, relevant to theoretical analysis or to practical applications, will be considered in a later paragraph.

Figures 1-3 show plots concerning the mean activity coefficient at the experimental level of representation (Lewis-Randall, LR frame [6, 8]), and the same is for the osmotic coefficient (Figure 4), which is obtained by integration of the Gibbs-Duhem Equation [6] after calculating $\gamma_\pm$. It is here outlined that comparison

will not be reported with results available only at the Mc Millan-Mayer level of representation (MM frame [6-8]). Conversion from LR to MM is formally possible, as known [19-22], but rather complicated in practice unless relatively simple models are adopted [23]. Approximate LR<->MM converting equations are to use [21] that are moderately reliable when medium-highly concentrated solutions are investigated. So, despite their historical or conceptual importance for theoretical understanding of ionic solutions, reported results from classical approaches, that are either limited to the MM frame or do not clearly refer to practical cases, will not be considered in the present work.

As a further, important, remark, it is here to say that QRL at present depends on one adjustable parameter (at given pressure $P$ and temperature $T$). Such a parameter is experimentally known, to note, for many common salts either symmetric or asymmetric [1-5]. The QRL parameter corresponds to a well-defined concentration, called $c_{lim}$ ($c_{lim}(P, T)$ on molar scale, mol/dm$^3$, $m_{lim}(P, T)$ on molal scale, mol/kg; see later), that also gives the upper concentration limit of applicability of the model (at given $T$ and $P$). Experimental $c_{lim}$ values, for aqueous electrolytes at various $T$ and $P$, range from 1 M to 8 M (about). For systems for which $c_{lim}$ cannot be experimentally reached, then the whole available concentration range (up to saturation or even super-saturation) can be considered by taking $c_{lim}$ as a tunable parameter.

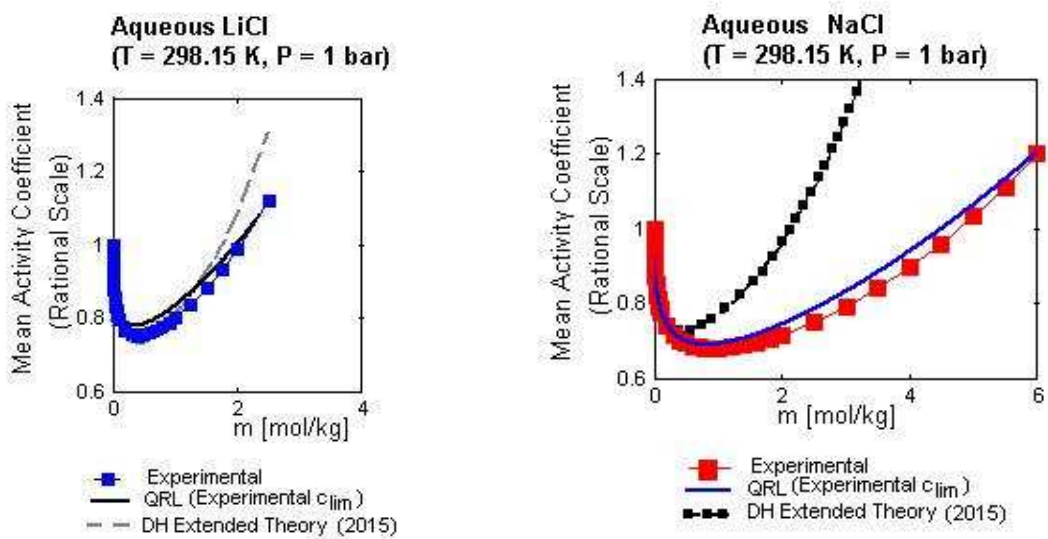

**Figure 1     Simplified Approaches to Ionic Solutions. Single-Parameter Models.** The mean activity coefficient is converted from molal to rational scale for a comparison with results given in Ref. [24]. For the cases presented in Figure 1, both QRL and DH Extended Theory [24] only use fully experimental information. In the DH Extended Theory, ionic radii are set to their Pauling values [24]; with QRL, experimental $c_{lim}$ values are 2.26 mol/dm$^3$ for LiCl, and 5.38 mol/dm$^3$ for NaCl (experimental data in Ref. [25]).

From Figure 1 it is evident that results from the DH Extended Theory [24] can largely diverge from experimental data yet at moderate concentrations. This outlines a general difficulty encountered by most models when tested versus fully experimental information. To solve this kind of problems, a typical procedure is introducing additional parameters, in theory, endowed with some physical-chemical definition or meaning, in practice, often quantified through numerical fittings and data-regression techniques. Not surprisingly, accuracy can remarkably be increased by increasing parameterisation, as shown by the example in Figure 2 below.

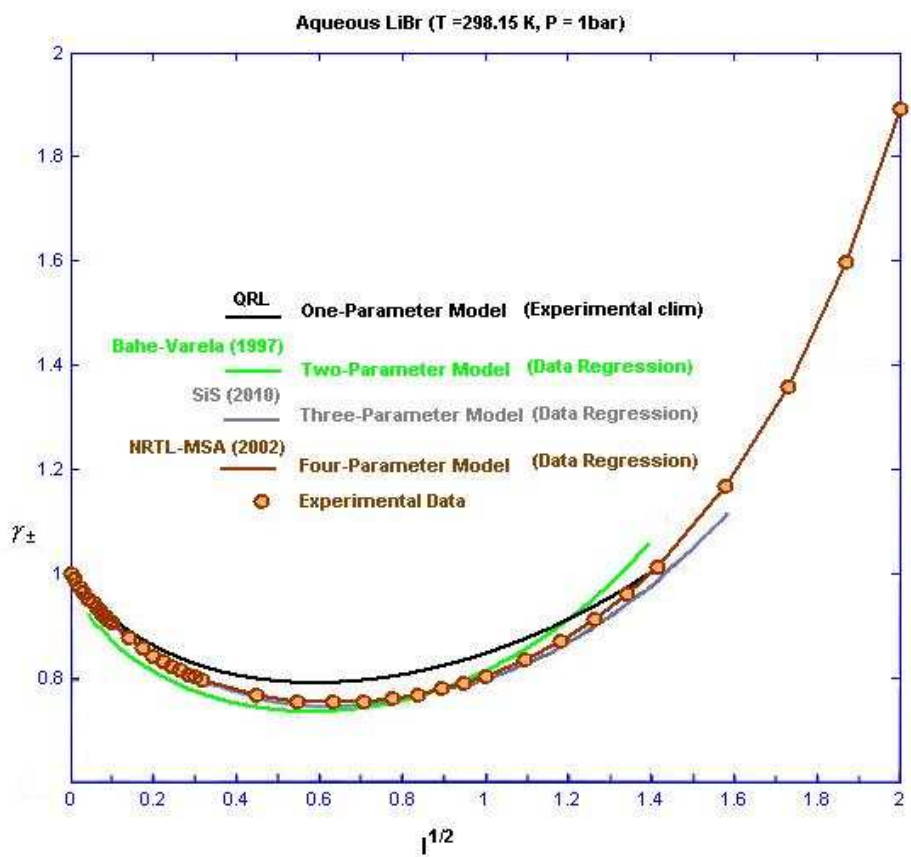

**Figure 2 Comparison among Multi-Parameter Models for Electrolyte Solutions** Abscissas are given in terms of square root of Ionic Strength I (molal scale) [6], for sake of comparison with results given in [14]. QRL: experimental $c_{lim}$= 1.855 mol/dm$^3$ [25]. Bahe-Varela [16]: the indicated parameterisation should be the minimal required by the approach. Single-ion Shell (SiS) [14]. Non-Random-Two-Liquid Model + Mean Spherical Approximation (NRTL-MSA) [26]: in Ref. [26] results for LiBr were given up to about 20 mol/kg, not wholly reported here for the sake of readability.

From Figure 2, it is also evident that, at least in principle, wider concentration-ranges can be considered when using a higher number of fitting parameters. However, concerning the concentration range that models can satisfactorily investigate, there is, in general, no simple rule to apply: upper concentrations are usually decided after best-fitting procedures from experimental data. This is not the case of QRL, where, to note, *the limit upper concentration, $c_{lim}$, is defined as the (upper) concentration corresponding to $\gamma_\pm$ =1*. As already said, for aqueous solutions $c_{lim}$ is usually found in the medium-high concentration range. However QRL has not yet been developed to investigate concentrations higher than $c_{lim}$.

Concerning accuracy, QRL performance can be competitive even with many-parameter models, as shown by the examples in Figure 3 below, and this is a significant aspect, if one considers that all the illustrated cases did not imply any best-fitting by QRL, since fully experimental information was available for them.

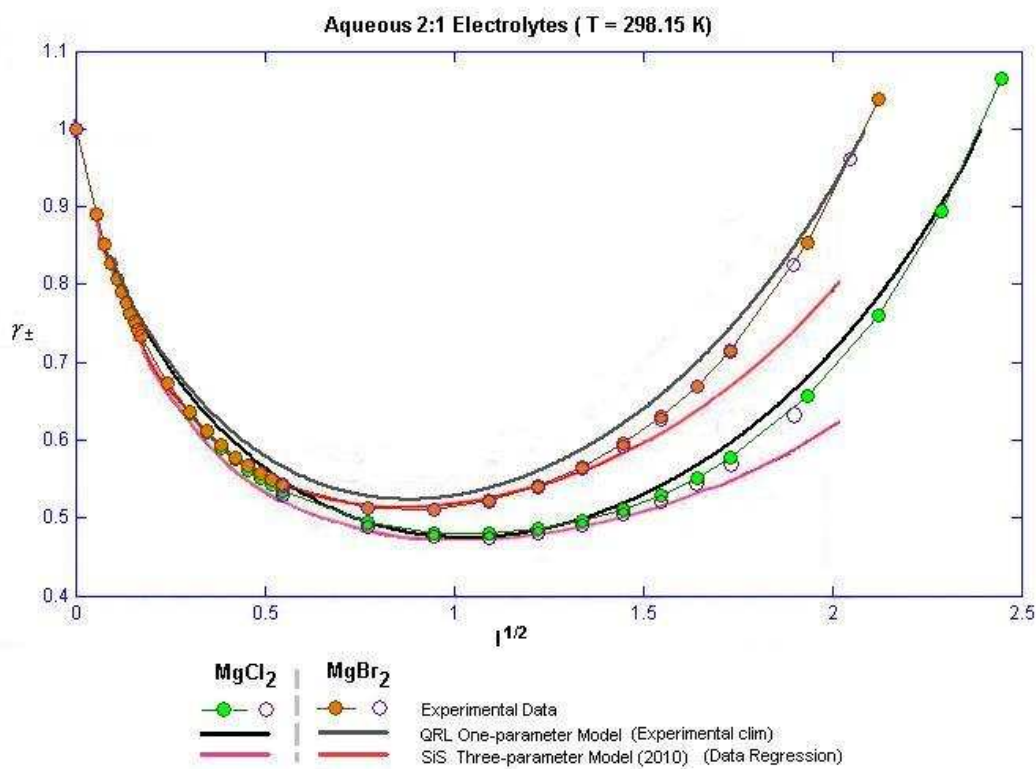

**Figure 3 Comparison between QRL and SiS** Abscissas are given in terms of square root of Ionic Strength I (molal scale) [6], for sake of comparison with SiS results [14]. For QRL, experimental $c_{lim}$ values are 1.846 mol/dm$^3$ for MgCl$_2$, and 1.377 mol/dm$^3$ for MgBr$_2$. Experimental data in Ref. [25] (filled and empty circles). With SiS, three adjustable parameters are used.

An important issue encountered by many-parameter theories is the agreement between formal definitions of parameters and their numerical values. The most classical parameterisation is concerned with ion-size and "closest-approach" distances [6-8], and there exists a plethora of interpretations based on radial distribution functions, solvation-shell, ion association or other concepts of Solution Theory [6-8]. Further parameterisation approaches include interaction-energy parameters (e.g., in local composition models [8, 15]), and dielectric-permittivity parameters [6-8, 13, 27] (either dielectric permittivity of solution, or dielectric permittivity of solvent as modified by the presence of solute). Although rigorous in principle or partially supported by measurements, huge parameterisation generally suffers from lack of experimental confirmation and from controversial results. Exemplifying cases are given by various generalizations of the Mean Spherical Approximation (MSA). In Ref. [28], a three-parameter fitting procedure (with respect to experimental $\gamma_\pm$), based on a MSA model, yielded unphysical results (negative radii) for some salts; in BIMSA (Binding MSA [13]) a parameter was included to account for dielectric-permittivity effects, however, some disagreement between calculated and measured permittivities forced redefining the permittivity parameter at a microscopic level [13, 29].

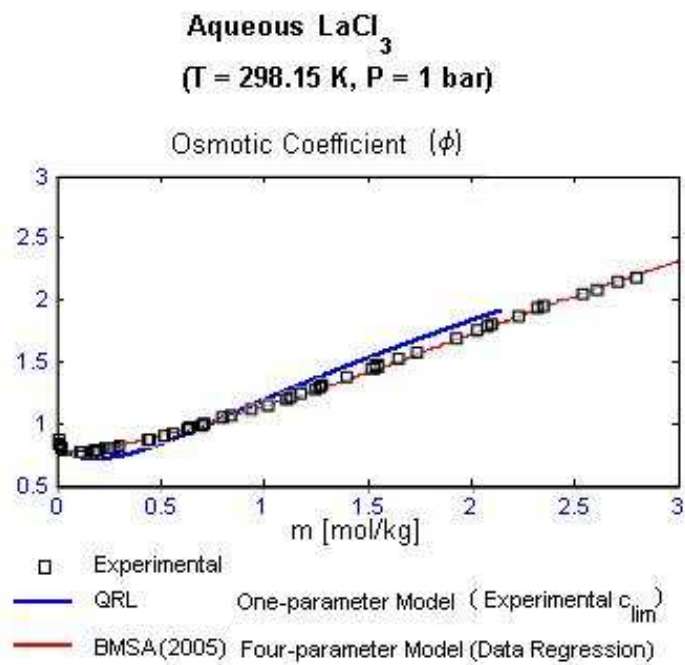

**Figure 4 Comparison between QRL and BIMSA** In BIMSA [13], four parameters (at given $P$ and $T$) are numerically calculated by means of data-regression. With QRL, $c_{lim}$ = 1.99 mol/dm$^3$ for LaCl$_3$ (experimental data in [25, 30, 31]; the corresponding molal value, $m_{lim}$, is 2.14 mol/kg).

In the following paragraphs it will be seen that, although belonging to the class of simplified approaches, QRL can provide very interesting results since its simple parameterisation is more significant, from a theoretical point of view, than so far intuited or suspected: this will particularly be outlined later on.

In what follows, a general overview of the QRL theory will first be presented. Then, some preliminary results will be discussed, in particular concerned with volumetric and thermal properties of electrolyte solutions (under current investigation, unpublished work).

**THEORY**

*Pseudo-Lattice Approach*

The first idea in QRL [1] is represented by the use of an "ionic-lattice" frame in place of the classical "central-ion" frame, the latter proposed by Debye and Hückel [6] for describing the statistical behaviour of ions and molecules in a solution. The ionic-lattice frame is composed by: (1) a fixed set of space points; (2) a fixed set of hypothetical (reference) charges.

The space-point set (1) allows for considering the solution volume conceptually divided into cells, and it can be set according to the electrolyte space-group geometry in the crystalline state. However, for sake of simplicity and without loss of generality, electrolytes of the same order will hereinafter refer to a same space group, e.g., the space group n. 221 (Cubic-System [32]) will hereinafter be used for the whole 3:1 electrolyte class[1]. Note that the lattice space-frame just provides a convenient set of space points from which observing, and statistically measuring, what is happening in the solution: in the DH Theory, this task is performed by the (mobile) space-frame centred on the reference ion [6-8].

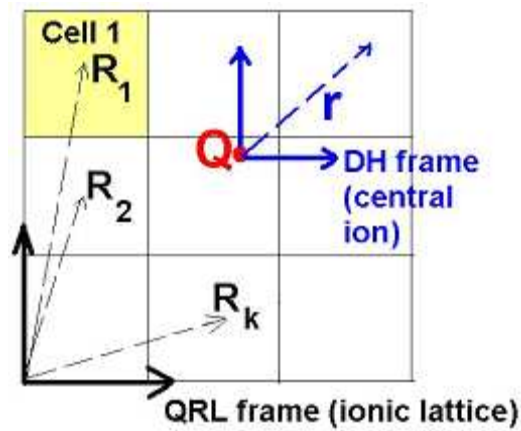

**Figure 5 QRL Approach. Ionic-Lattice Space-Frame.** Pictorial representation of a ionic-lattice space-frame (plane-view grid), where cells are individuated by a reference set of lattice points (**R**$_1$,..**R**$_k$,..). For comparison, the Debye-Hückel (DH) space–frame, centred on the "central ion" Q, is also represented.

Concerning the reference charges (2), they are here said hypothetical in that they are those expected if solute ions were located within the lattice cells according to the ionic-crystal configuration, so as to provide, on their whole, the so-called "reference electrostatic configuration". But, of course, in reality solute ions will not generally coincide with reference charges, since they will likely be found largely far or widely moving with respect to their lattice positions: while, in the DH picture, the reference charge is that of the solute-ion that is located at the reference space position (mobile "central–ion" frame), conversely, in the QRL picture, reference charges are only those hypothetically expected at the reticular points (fixed "ionic-lattice" frame) in the presence of electrostatic interactions only.

It is assumed that, on time and space-average, the net charge interesting a given cell will amount to the bare charge of a solute ion, and net charges will be distributed in cells according to the electrolyte-crystal scheme. This assumption is advantageous because the electro-neutrality condition [7] will be easier to satisfy. However, it is stressed out again that, by no means, the net charge interesting a cell can be attributed to a given (or "central") ion, rather, there will be a population of solute ions and solvent molecules that will be observed "fluctuating", with time, in that cell.

*Carriers and Effective Carriers*

It can be objected that assuming net charges located in cells according to the ionic-crystal configuration does not avoid, after all, to see somewhat a crystal-like structure in the solution, even at strong dilution. The answer comes from the second idea in QRL [1], that is, ions and molecules fluctuating in a given cell,

---

[1] Note that, for example, crystal structures of Rare-Earth Chlorides usually belong to the Space Group n. 176 (P6$_3$/m,

observed along with time like in a sequence of photographs, behave as if they were almost-instantaneous groupings, generally called "carriers of charge", with changing size and / or composition: on the whole, a crystal-like structure widely distorted and continuously modified. The idea is not completely new in pseudo-lattice approaches [6, 34, 35], and indeed, disordered lattices to represent ionic solutions were considered, since 1918, by Ghosh (see Ref. [6] for a historical overview). What is a novelty introduced by QRL is the use of flexible carriers rather than "hard-sphere" ions. The overall picture is easier to visualize at strong dilution, where carriers correspond to solute ions that, although endowed with a finite-size, can be treated as point charges (as done by literature models [6-8]), the effect of their physical extent being negligible at strong dilution. However, point-charge ions can largely move in the solution because thermal energy is definitely predominant over their electrostatic interactions [1, 5]. The resulting effect is that, on time and space average, carriers interesting a given cell "reproduce" a sort of density of charge that widely spreads out from the cell. By statistically measuring the extent of this charge density from the reference lattice point of the cell, assuming that its spatial integration will give a net charge according to the position of that cell in the ionic crystal, then the so-called "effective carrier", associated with the given cell, will be "visualized". Actually, effective carriers, rather than carriers, are used to derive the main equations in QRL. So, an effective carrier is the statistical result after averaging (on time and space) over carriers and their movements. An effective carrier accounts for flexible carrier size and composition, and for stochastic movements in the solution due to Brownian Motion in a thermal bath [1, 5]. An effective carrier is modelled by means of a charge density, which is approximately assumed Gaussian-distributed [1] with centre on the reference lattice point of the cell. The linear standard deviation (l.s.d.), called $U$, of such a Gaussian distribution accounts for the physical extent of carriers combined with the extent of their displacements. It is clear that $U$ is very large, and ideally approaches infinity at infinite dilution [1, 5].

An important point is the physical meaning of carriers when the solute concentration increases. As above said, a carrier generally represents a "grouping" of ions and molecules that are concerning a cell at a given (time and space) observation. Although very general and not restrictive, this definition does not explain too much about the carrier nature. The exact microscopic composition of a carrier is not yet available by QRL, however some considerations can be helpful. First, concerning the solvent molecules belonging to a carrier, one can refer to those molecules that appear "perturbed" with respect to their "bulk" state because of the presence of solute ions (of course, the ion-solvation molecules are included). The bulk state is that of the solvent in absence of solute, which can be modelled by a continuous-medium, or unstructured solvent (according to literature [6]). Note, solvent molecules included in a carrier may, partially or fully, reacquire their bulk state after a while. However, the higher the concentration the higher the fraction of solvent molecules included in a carrier. Second, concerning the solute ions belonging to a carrier, when

---

Hexagonal System [32, 33]).

more ions will be found within a same cell, there will also be more cells affected by a same carrier. In this sense, the extent of a carrier increases with increasing concentration, also considering that cell volumes decrease in the meanwhile, and more intense interactions are allowed among ions becoming closer (note, this also allows for ion association, according to literature [6]). If solvent molecules belong to a carrier depending on their degree of perturbation state with respect to their bulk state, then solute ions belong to a carrier depending on the strength of their interactions with respect to their "almost free-particle state" (at infinite dilution). When all cells are to be included in all carriers, then carriers have reached their maximum extent, which is ideally infinite: at such an extreme condition, occurring at the concentration $c_{\lim}$, the charge density, modelling an effective carrier, will be a Gaussian distribution where, again, the l.s.d. $U$ will approach infinity (and $\gamma_\pm$ approaches 1). At present, this is also the limit condition for the applicability of QRL (the model so far only deals with $\ln(\gamma_\pm)\leq 0$), developed on mesoscopic-scale (no microscopic scale was so far investigated). A mesoscopic approach allows for using tools from Continuum Electrostatics.

We hereinafter suppose that in a volume V there are N solute molecules giving rise to $\nu_+$ N carriers with charge $Q_+$, and $\nu_-$ N carriers with charge $Q_-$. We also put:

$$\nu = \nu_+ + \nu_-; \qquad \rho = N/V; \quad R = 1/(\nu\rho)^{1/3} \tag{1}$$

In Eq. (1) $R$ indicates the mean inter-ionic distance. In practical applications, one puts V = 1 m$^3$, so that $N=10^3 N_{Av} c$, where $N_{Av}$ is the Avogadro number, and $c$ the molar concentration [mol/dm$^3$]. Just to remind here that concentration scales can easily be converted from each other by means of standard formulae.

$$c = \frac{10^3 md}{10^3 + m W_s} \tag{2}$$

In Eq. (2), $m$ is the molal concentration of solute [mol/kg], $d$ is the absolute solution density (g/cm$^3$), and $W_s$ is the molecular weight of the solute. Hereinafter, $Q_+ = z^+ Q$ and $Q_- = z^- Q$, where $z^+$ and $z^-$ are the ion valences, while Q indicates the (absolute) electron charge = 1.66x10$^{-19}$ C. The electro-neutrality condition is given by:

$$\nu_+ Q_+ + \nu_- Q_- = 0 \tag{3}$$

For a reference lattice frame with reference points (lattice vertices) $\{\mathbf{R}_A\}$ and net charges $\{Q_A\}$ (where either $Q_A=Q_+$ or $Q_A=Q_-$), the effective carrier A is described by the charge density below [1]:

$$\Omega_A(\mathbf{r}) = \frac{Q_A}{\left(\sqrt{2\pi} U\right)^3} \exp\left(-\frac{|\mathbf{r} - \mathbf{R}_A|^2}{2U^2}\right) \tag{4}$$

In Eq. (4) $\mathbf{r}$ is a generic position with respect to a fixed origin in the space frame. Given the $\{\mathbf{R}_A\}$ set, full knowledge of $\Omega_A(\mathbf{r})$ requires an expression for the l.s.d. $U$. Discussion and formulae for calculating $U$ will be subject of later paragraphs.

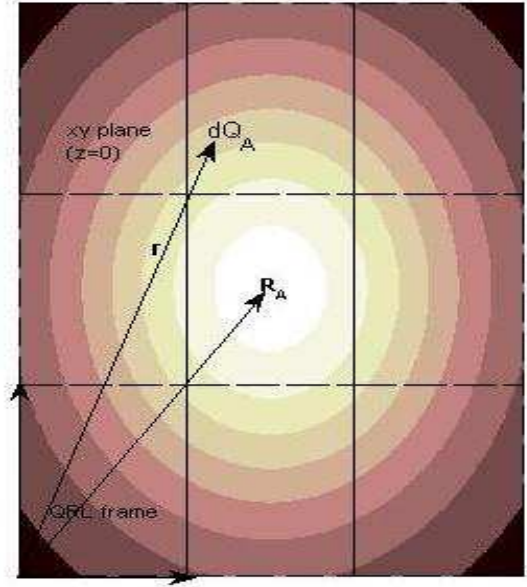

**Figure 6. QRL Approach. Effective Carrier** An effective carrier, with overall charge $Q_A$, is pictorially shown, referring to a cell of a FCC (Face Centred Cubic) ionic lattice, which is a typical crystal structure for aqueous 1:1 electrolytes at ambient conditions [32]. The corresponding lattice point $\mathbf{R}_A$ belongs to the z=0 plane. Black grid: plane view of the reference lattice. In the QRL space-frame, an infinitesimal charge amount $dQ_A$, belonging to the effective carrier, is shown at position $\mathbf{r}$. In the reference (hypothetical) electrostatic configuration, the ionic charge $Q_A$ is located at $\mathbf{R}_A$.

*Mean Activity Coefficient*

Note. In this paragraph, $U$ (Eq. (4)) will be let indicated. Detailed discussion about $U$ will be done later on.

The third idea of QRL [1] is calculating $\ln(\gamma_\pm)$ starting from calculating the mean energy $E_A$ of an effective carrier A, due to its interactions with the other effective carriers. This idea allows for avoiding explicit knowledge of the microscopic nature of carriers. Calculations, reported in details in Refs. [1, 3], first required solving a Poisson equation to find the potentials $\{V_B\}$ generated by the set of effective densities $\{\Omega_B\}$ according to Eq. (4), yielding:

$$V_B(\mathbf{r}) = \frac{Q_B \, \mathrm{erf}\left(\dfrac{|\mathbf{r}-\mathbf{R}_B|}{\sqrt{2}U}\right)}{4\pi\varepsilon|\mathbf{r}-\mathbf{R}_B|} \qquad (5)$$

In Eq. (5), the symbol $\varepsilon$ indicates the dielectric permettivity of solvent; the symbol *erf* indicates the Error function. Then, the overall contribution from any B, excluding A, was space-averaged with respect to $\Omega_A/Q_A$ (the rate expresses a density probability) leading to:

$$E_A = \frac{Q_A}{4\pi\varepsilon} \sum_{\substack{B \\ B \neq A}} \frac{Q_B}{R_{AB}} \, \mathrm{erf}\left(\frac{R_{AB}}{2U}\right) \qquad (6)$$

In Eq. (6), $R_{AB} = |\mathbf{R}_A - \mathbf{R}_B|$. The QRL formula for $\ln(\gamma_\pm)$ followed from the relationship [4]:

$$\ln(\gamma_\pm) = \frac{1}{2k_BT} \sum_A E_A / (\nu_+ N + \nu_- N) \qquad (7)$$

In Eq. (7), the ½ factor avoids interactions counted twice, as in the so-called "Robinson-Stokes charging process" [6]. Finally:

$$\ln(\gamma_\pm) = \frac{1}{k_BT} \frac{1}{2\nu N} \frac{1}{4\pi\varepsilon} \sum_{\substack{A,B \\ A \neq B}} \frac{Q_A Q_B}{R_{AB}} \text{erf}\left(\frac{R_{AB}}{2U}\right) \qquad (8)$$

Note that Eq. (8) can also be cast into single-ion activity coefficients $\gamma_+$ and $\gamma_-$, after separately evaluating $E_{A+}$ and $E_{A-}$. This theoretical aspect and its implications will not further be considered in the present work, but will be subject of future research. Eq. (8) is useful for theoretical investigations [4], while fast numerical computation of $\ln(\gamma_\pm)$ was proposed via integral equations containing elliptic functions [1-4]. Helpful manipulations of lattice sums were also performed according to the space-group geometry. Below, some formulae are summarised referring to simplest lattice geometries and space groups (all belonging to the Cubic System [32]) for symmetric, uni-divalent and uni-trivalent electrolytes. Application and generalization of formulae to higher-order electrolytes is of course possible, but will not be considered in the present work.

1) *Symmetric electrolytes* [1-3]. $Q_+ = -Q_- = Q$; $\nu = 2$. Space Group n. 225, Fm3m (NaCl-like) [32]. A generic lattice point A is represented by $\mathbf{R}_A = (-1)^{n+m+p} R$, where (n, m, p) will hereinafter indicate any arbitrary triplet of integers.

$$\ln(\gamma_\pm) = \frac{Q^2}{2\sqrt{2\pi}^3 \varepsilon k_B T U} \int_0^{\frac{1}{\sqrt{2}}} \left\{ \left( \sum_{n=-\infty}^{n=+\infty} (-1)^n e^{-R^2 \frac{n^2 u^2}{2U^2}} \right)^3 - 1 \right\} du = \frac{Q^2}{2\sqrt{2\pi}^3 \varepsilon k_B T U} \int_0^{\frac{1}{\sqrt{2}}} \left\{ \theta_4\left(0, e^{-\frac{R^2 u^2}{2U^2}}\right)^3 - 1 \right\} du \qquad (9)$$

In Eq. (9), $k_B$ is the Boltzmann Constant, and $\theta_4(\xi, z) \equiv \sum_{n=-\infty}^{n=+\infty} (-1)^n e^{2jn\xi} z^{n^2}$ is the Fourth Jacobi Theta Function [36], with $\xi = 0$ and $z = e^{-\frac{R^2 u^2}{2U^2}}$.

2) *Uni-divalent electrolytes* [1-3]. $Q_+ = 2Q$, $Q_- = -Q$; $\nu = 3$. Space Group n. 225, Fm3m (CaF$_2$-like [32]).

$$\mathbf{R}_{A+}: \sqrt[3]{\frac{3}{2}} R(2n,2m,2p), \sqrt[3]{\frac{3}{2}} R(2n+1,2m,2p), \sqrt[3]{\frac{2}{3}} R(2n,2m+1,2p), \sqrt[3]{\frac{3}{2}} R(2n,2m,2p+1); \mathbf{R}_{A-}: \sqrt[3]{\frac{3}{2}} R(n+1/2,m+1/2,p+1/2) \qquad (10)$$

$$\ln(\gamma_\pm) = \frac{1}{2k_BT} \frac{1}{\sqrt{2\pi}^3} \frac{2Q^2}{3\varepsilon} \frac{1}{U} \int_0^{\frac{1}{\sqrt{2}}} \left\{ 2(A^3(u) - 1) + 6A(u)B^2(u) + (A(u) + B(u))^3 - 1 - 2C^3(u) \right\} du \qquad (11)$$

In Eq. (11), the following auxiliary functions $A(u)$, $B(u)$ and $C(u)$ appear.

$$A(u) = \sum_{n=-\infty}^{\infty} e^{-4n^2 \left(\frac{3}{2}\right)^{\frac{2}{3}} \frac{R^2 u^2}{2u^2}} \qquad B(u) = \sum_{n=-\infty}^{\infty} e^{-(2n+1)^2 \left(\frac{3}{2}\right)^{\frac{2}{3}} \frac{R^2 u^2}{2U^2}} \qquad C(u) = \sum_{n=-\infty}^{\infty} e^{-(n+1/2)^2 \left(\frac{3}{2}\right)^{\frac{2}{3}} \frac{R^2 u^2}{2U^2}} \qquad (12)$$

In Eq. (12), the Third Jacobi Theta Function [36], $\theta_3(\xi, z) \equiv \sum_{n=-\infty}^{n=+\infty} e^{j2n\xi} z^{n^2}$, might also be evidenced (for further details upon elliptic functions, see, e.g., Refs. [36,37]). Note that, in Eq. (12), the l.s.d. $U$ is assumed to be the same for positive and negative effective-carriers, according to considerations done in Ref. [4] and differently from what done in Refs. [1-3].

3) *Uni-trivalent electrolytes.* $Q_+=3Q$, $Q_-=-Q$, $\nu=4$. Space Group n. 221 (Cubic System).
$\mathbf{R}_{A+}: R(2n,2m,2p)$; $\mathbf{R}_{A-}: R(2n+1, 2m, 2p)$, $R(2n, 2m+1, 2p)$, $R(2n, 2m, 2p+1)$ (13)

$$\ln(\gamma_\pm) = \frac{1}{2k_B T} \frac{1}{\sqrt{2\pi}^3} \frac{3Q^2}{4\varepsilon} \frac{1}{U} \int_0^{\frac{1}{\sqrt{2}}} \left\{ 4\left(A^3(u) - 1\right) - 6A^2(u)B(u) + 2A(u)B^2(u) \right\} du \qquad (14)$$

In Eq. (14), the auxiliary functions are:

$$A(u) = \sum_{n=-\infty}^{\infty} e^{-(2n)^2 \frac{R^2 u^2}{2u^2}} \quad ; \qquad B(u) = \sum_{n=-\infty}^{\infty} e^{-(2n-1)^2 \frac{R^2 u^2}{2u^2}} \qquad (15)$$

Note that, in Ref. [4], Rare-Earth Chlorides and Bromides were investigated setting the reference lattice according to the Space Group n. 176, (P6$_3$/m, Hexagonal System [32, 33]). However, calculations were rather long and complicated, so they will no longer reported here. Simulation results in Figure 4, concerning aqueous LaCl$_3$ at ambient conditions, have been performed according to Eqs. (13-15).

Formulae above for calculating $\ln(\gamma_\pm)$ can be applied provided that an expression for $U$ is given. This will be done in the following paragraphs.

*Infinite Dilution Limit*

The Pseudolattice Theory [6, 34, 35] was presented in literature at various levels of maturity, yielding modern models that are generally applied to highly concentrated solutions and ionic liquids [35]. The main motivation is that an ionic lattice can provide some configuration pattern for local structures, and this hypothesis is obviously more reasonable (and also supported by experimental information [38]) with high levels of solute concentration. However, another important motivation arises from the theoretical drawback represented by the so-called "cube-root law" (with respect to concentration), that reminds to some crystal-like behaviour and that appears in formulae for $\ln(\gamma_\pm)$ given by all pseudo-lattice models [6, 16, 34] but QRL [1]. As known, a cube-root law is not consistent with the Debye-Hückel Limiting Law (DHLL) [6]. Attempts to derive DHLL from cube-root laws were proposed during decades (in this connection, see, e.g., Refs. [6, 39, 40]), but without success. A main outcome of QRL is its convergence to DHLL, which was shown in Ref. [5]. The key step was the derivation of the limiting formula below (where $U_\infty$ stands for $U$ at the infinite-dilution).

$$U = \frac{U_D}{\sqrt{\pi}} \equiv U_\infty \quad (\rho \to 0) \qquad \text{(Infinite-Dilution Limit)} \qquad (16)$$

In Eq. (16) the DH Screening Length ($U_D$) is evidenced.

$$U_D = (\varepsilon k_B T)^{1/2}/(\nu\rho|Q_+Q_-|)^{1/2} \qquad (17)$$

In Ref. [5], it was shown that DHLL can be obtained from QRL using $U = U_\infty$ at the infinite dilution limit.

$$\ln \gamma_\pm \underset{\rho \to 0}{\approx} \frac{1}{k_B T} \frac{Q_-Q_+}{8\pi\varepsilon} \frac{1}{\sqrt{\pi}U} \quad \text{(QRL)} \quad = \frac{1}{k_B T} \frac{Q_-Q_+}{8\pi\varepsilon} \frac{1}{\sqrt{\pi}U_\infty} = \frac{1}{k_B T} \frac{Q_-Q_+}{8\pi\varepsilon} \frac{1}{U_D} \quad \text{(DHLL)} \qquad (18)$$

In Eq. (18), the first limit expression for $\ln(\gamma_\pm)$ arises from evaluating the QRL master equation (8) in the limit $\rho \to 0$ [5]. Equations (16, 18) are important from a theoretical point of view because they were derived independently from the DH Theory. Eq. (16) was derived after considering that, at very strong dilution, the reference electrostatic configuration (the ionic lattice) is "randomised" twice: 1) from the presence of Brownian Motion in a thermal bath, as generally supposed by QRL; 2) from the presence of too few ions that tend to behave as if they were almost-free particles such that no ionic lattice configuration can be obtained for a sufficiently long observation time. That is, as far as $\rho \to 0$, the electrostatic reference configuration becomes closer to an almost-randomised lattice were ions are located at vertices almost regardless of their charges. The theoretical importance of a pseudo-lattice derivation of the exact DHLL (proven first within the context of Fluctuation Solution Theory, by Kirkwood and co-workers [41, 42]) can also be realized by the following considerations. Most part of simplified approaches, in particular those born within the Primitive Model (PM) context [6-8], can be seen, after all, as generalized forms of the DH Theory, with respect to which they are, *de facto*, extended theories [43]. This also holds true for Pitzer Theory [17], Specific-Ion Interaction [6] and Dressed-Ion Theory [44]. In other words, theoretical consistency at infinite dilution is often forced, first, by starting from DHLL, second, by adding terms or by modifying parameters such that will be active only at non-infinitesimal concentrations. Note, forced (*a priori*) DHLL convergence also appears in lattice-based extensions of the PM (see, e.g., LRPM [45][2]). An important remark is that the independence from the DH Theory that characterizes QRL also allowed [5] satisfying the electro-neutrality condition (or "Zero-Order Moment-Equation"[7]) at the infinite dilution limit without introducing any "ad-hoc" parameterisation. However, in this connection, it must also be said that generalization to non-infinitesimal solute concentrations has not yet been done with QRL: this will be object of future research.

*From Infinite Dilution to Strong Dilution*

It was first suggested in Ref. [1] that $U$ should be of the order of $\frac{\sqrt{2}U_D}{\sqrt{\pi\alpha}}$ in the strong-dilution range, where the symbol $\alpha$ indicates the Madelung Constant [32] of the reference ionic lattice[3]. Then, in Ref. [5], it was more precisely proven that $U$ converges to $\frac{U_D}{\sqrt{\pi}}$ at the infinite dilution limit (Eq. (16)). It is important to note that no Madelung constant appears at the infinite dilution limit, according to the fact that lattice geometries must lose their specificity as $\rho$ approaches 0. In practical situations and with realistic concentrations, neglecting the infinite-dilute limit and using $U=\frac{\sqrt{2}U_D}{\sqrt{\pi\alpha}}$ *tout court* for very dilute solutions does not introduce appreciable differences on $\gamma_\pm$ (also considering that $\alpha$ is usually found not greatly different from 2, for example, $\alpha \approx 1.7476$ for FCC symmetric systems [32]). However, for sake of theoretical consistency and in order to compact formulae from infinite to strong dilution, the following unified expression was suggested [5].

$$U=\frac{U_D}{\sqrt{\pi}}\left[1+\frac{e^k c}{2k^2 c_{\lim}}\left(\sqrt{\frac{2}{\alpha}}(6k^2+1)-6k^2\right)-\frac{e^{2k}c^2}{k^2 c_{\lim}^2}\left(\sqrt{\frac{2}{\alpha}}(3k^2+1)-3k^2\right)+\frac{e^{3k}c^3}{2k^2 c_{\lim}^3}\left(\sqrt{\frac{2}{\alpha}}(2k^2+1)-2k^2\right)\right]$$

$$0 \leq c \leq c_{\lim}\exp(-k) \tag{19}$$

In the equation above, note that the exact equality $U=\sqrt{\frac{2}{\pi\alpha}}U_D$ is obtained at $c_0 \equiv c_{\lim}\exp(-k)$, where $c_0$ is assumed to be a representative concentration of the electrolyte behaviour in the strong-dilution range [1]. In Eq. (19), the constant $k$ also appears, $1<k<\infty$, in detail described in previous works [1, 3] and here reconsidered. Because of the meaning of $c_0$, $k$ must be sufficiently large such that $c_0$ will be sufficiently small so as to lie within the strong-dilution range. Determination of $k$ is a crucial point since it also involves investigation of non-dilute solutions, as we will see in the next paragraph. A first way to determine $k$ was a semi-empirical procedure [1], for sake of simplicity here summarized upon 1:1 aqueous electrolytes at ambient conditions. For these cases, a practical strong-dilution range should not exceed $10^{-3} \div 10^{-2}$ M, according to experimental literature [25]. This suggested [1] taking $k$ of the order of $10 \div 11$ about [4], such that $c_0 \leq 5\times 10^{-4}$M even considering $c_{\lim}$ as large as 20 M [5]. In the range $[0, c_0]$, although $c_0$ rigorously depends on $c_{\lim}$ (that is, $c_0$ is salt specific), when using $U$ of Eq. (19), numerical differences on $\gamma_\pm$ among different salts (of the same electrolyte order) are *de facto* vanishing, according to classical DH results [6]. However, it must be remembered that the unified expression for $U$ given in Eq. (19) is just an approximation based on few power-expansion terms of $U$ with respect to $c/c_{lim}$ (see Ref. [5] for details), and other approximations might also be adopted. So, for sake of simplicity and according to what done in previous works [1-4], the expression $\sqrt{\frac{2}{\pi\alpha}}U_D$ will hereinafter be assumed as the representative linear standard deviation concerning effective carriers at strong dilution.

---

[2] Concerning LRPM, the approach also seems quite limited in terms of its applicability to practical cases [45].
[3] With QRL, Madelung constants refer to $R$ (mean inter-ionic distance, see Eq. (1)) and not to the minimum distance between ions of opposite charge [32]. However, in Refs [1-3] Madelung constants were also reported with respect to the minimum distance between two ions of opposite sign, according to the most common definition of these constants in literature [32]. Symbols used for indicating Madelung constants are described within each reference.
[4] Higher $k$ values were found suitable for asymmetric electrolytes [1-4].

*From Strong Dilution to Medium-Low Dilution*

For going from strong-dilution to medium-low dilution the following expression for $U$ was derived [1]:

$$U = \frac{U_D}{\exp(1)\sqrt{\frac{\pi\alpha}{2}}\frac{1}{k}\ln\left(\frac{c_{lim}}{c}\right)\left(\frac{c}{c_{lim}}\right)^{\frac{1}{k}}}, \quad c_{lim}\exp(-k) \leq c \leq c_{lim}, \quad 1<k<\infty \tag{20}$$

Note, Eq. (20) was rigorously obtained for concentrations close to $c_{lim}$, and then extended for continuity over the whole range $[c_0, c_{lim}]$, so it should be considered as an approximation (however satisfactory!) in the medium-low concentration range. On their whole, Eq. (19) and Eq. (20) give expressions for $U$ over the concentration range $[0, c_{lim}]$. The continuity condition between Eq. (19) and Eq. (20) is imposed at $c=c_0$, which, to note, minimizes $U$. In theory, imposing continuity at $c=c_0$ between different expressions of $U$, each for a different concentration range and for a different electrolyte behaviour, points out a formal transition from strong to lower dilution after $U$ has reached its minimum value. In practice, Eq. (20) moderately changes with respect to Eq. (19) until concentrations maintain sufficiently small (say, less than $10^{-3} \div 10^{-2}$) [6].

Eq. (20) was derived [1] after defining $1/k$ as the fraction of the total amount of carriers that, as $c \rightarrow c_{lim}$, still keep their "full" point-charge character, such that, first, their size is negligible, second, they can statistically be described by effective carriers whose linear standard deviation is $\sqrt{\frac{2}{\pi\alpha}}U_D$ (about). It might be objected that $U_D$ is no longer large at high concentrations (in other words, point-charge carriers are no longer widely fluctuating within the solution), however, since carriers keeping some negligible size are almost zero if $c \rightarrow c_{lim}$, then the fraction $1/k$ must refer to a vanishing total amount of them. $\sqrt{\frac{2}{\pi\alpha}}U_D$ can then be considered in some functional relationship with $U$, $k$ and $\frac{c}{c_{lim}}$, and put in the form:

---

[5] For aqueous electrolytes at ambient conditions, experimental $c_{lim}$ values usually range from 1 M to 10 M about [25].
[6] Relative changes of $U$ of Eq. (20) with respect to Eq. (19) are moderate up to $10^{-4} \div 10^{-3}$ M even if $k$ is as large as 20 (which is the case for 3:1 electrolytes) and $c_0$ is very small (of the order of $10^{-10}$ or less).

$$\sqrt{\frac{2}{\pi\alpha}}U_D = F\left(k, \frac{c}{c_{\lim}}, U\right) \qquad c \to c_{\lim} \tag{21}$$

Since $U \to \infty$ when $c \to c_{\lim}$ [7], then, from a first-order expansion [8] of $F$ with respect to $U$, one writes

$$\sqrt{\frac{2}{\pi\alpha}}U_D \approx f\left(k, \frac{c}{c_{\lim}}\right)U \qquad c \to c_{\lim} \tag{22}$$

Due to finiteness of $\sqrt{\frac{2}{\pi\alpha}}U_D$, it must be: $f\left(k, \frac{c}{c_{\lim}}\right) \to 0$ when $c \to c_{\lim}$. Further investigation finally yielded (see [3], Appendix A, for details):

$$f\left(k, \frac{c}{c_{\lim}}\right) \approx \frac{e^1 \ln\left(\frac{c_{\lim}}{c}\right)}{k\left(\frac{c}{c_{\lim}}\right)^{1/k}} \tag{23}$$

Eq. (22) and Eq. (23) straightforwardly give Eq. (20). Few additional comments are however important, concerning Eqs. (23). The term $\sqrt{\frac{2}{\pi\alpha}}U_D$ was also indicated [1] with $U_{I-I}$ in order to emphasize its origin from long-range ion-ion interactions (while considering the solvent as an unstructured medium). More generally, and according to previous discussion about the meaning of carriers and effective carriers, the linear standard deviation $U$ was also supposed to include some $U_{I-D}$ contribution [1] due to short-range ion-dipole interactions (that is, due to that part of solvent included within carriers)[9]. However, at the current State of Art, it is just possible speaking about $U_{I-I}$ and $U$ and their relationship [5], since investigation has not yet been performed concerning the microscopic nature of carriers. That said, from Eq. (20) it is evident that $U_{I-I}$ and $U$ differ from a factor $f\left(k, \frac{c}{c_{\lim}}\right)$ (Eq. (23)), which should be considered arising from the increasing size of carriers with concentration, and resulting from short-range interactions among ions and molecules. If the solute concentration increases, then $U_{I-I}$ becomes a smaller $f\left(k, \frac{c}{c_{\lim}}\right)$ "fraction" of $U$, stating, once again, that long-range (ion-ion) interactions are overcame by short-range interactions. Note, since Eq. (20) arises from extending from $c \to c_{\lim}$ down to $c = c_0$ [3], considerations here made should be taken in an approximate sense. In Ref. [3] it was suggested that, for dilute solutions only, there might be an additive rule such that $U$ would approximately be equal to $U_{I-I}$ +$U_{I-D}$. However, at present there is no QRL theory at microscopic scale, so suggestions reported in Ref. [3] (see Eqs. (8, 9) in the reference) are to consider purely indicative.

---

[7] Otherwise said, $c_{\lim}$ is the concentration at which carriers reach their maximum extent and effective carriers are described by the largest $U$.

[8] In Ref. [3], Appendix A, the functional $F$ was forgotten (see text concerning Eq. A11). The cited text erroneously refers to a first-order approximation of $U$, while it should refer to a first-order approximation of $F$ with respect to $U$.

[9] Plus, perhaps, other terms, not yet mentioned [1], such as $U_{D-D}$ that should account for contributions due to Dipole-Dipole interaction effects on carriers and effective carriers.

As anticipated in the previous paragraph, determination of $k$ is a fundamental point in the QRL approach. Besides semi-empirical procedures [1], the formula below was proposed in Ref. [3].

$$\frac{k}{e^1 \ln(k)} = \sqrt{\frac{\pi\alpha}{2}} \qquad 1<k<\infty \tag{24}$$

Eq. (24) is solved to obtain $k$ after $\alpha$ is given. For symmetric electrolytes and the FCC lattice, $\alpha = 1.7476$, so $k = 10.65$ (which is in excellent agreement with estimated $k$ (10÷11) with semi-empirical procedure [1]). Concerning asymmetric electrolytes, previous estimates of $k$ [1-4] still agree with Eq. (24), which yields $k = 14.54$ for uni-divalent electrolytes ($\alpha=1.679$) and $k = 19.69$ for uni-trivalent electrolytes ($\alpha = 2.39$). Calculations using space-group geometries other than those here presented [1-4] gave $k$ values of the same order of magnitude as those above. Using Eqs.(8, 19, 20, 24) allows for a single-parameter ($c_{lim}$) QRL model over the whole [0, $c_{lim}$] range at any given $P$ and $T$. However, derivation of Eq. (24) presents some issues that will be revised here below.

The main strategy to obtain Eq. (24) [3] was considering $k$ only depending on the reference ionic lattice through $\alpha$, that is, $k = k(\alpha)$. Discussion was then better driven by investigating some inverse relationship, $\alpha=\alpha(k)$. Since $k\geq 1$ by definition, the extrema $k=1$ and $k \to \infty$ were investigated. If $k=1$ then we should speak about an ionic lattice able to show the maximum opposition against changes upon its "point-charge" configuration. In such a case, the electrostatic energy should be highly predominant over the thermal energy even at very strong dilution (very large $R$), which definitely implies $\alpha \to \infty$ and $\sqrt{\frac{2}{\pi\alpha}}U_D \to 0$. As a consequence, if $k = 1$ then all carriers should keep their point-charge character regardless of their number, that is, $U \to U_{I-I} = \sqrt{\frac{2}{\pi\alpha}}U_D \to 0$ at any $c$ ($c_0 \leq c \leq c_{lim}$). Of course, there exists no ionic lattice endowed with $\alpha = \infty$, so the case $k=1$ is only a mathematical abstraction. On the other extreme, $k \to \infty$ means that no point-charge carriers are found when $c$ approaches $c_{lim}$, so the ionic lattice should be the most affected by any little perturbation, and $k \to \infty$ should provide the largest $U$. This can be obtained [3] from Eq. (20) by imposing $\frac{\sqrt{\alpha(k)}}{k} \to 0, k \to \infty$, which yields $U \to \infty$ at any $c$ ($c_0 \leq c \leq c_{lim}$).

In Ref. [3], investigation continued by studying the case $k \to \infty$ when $c \leq c_0$, which means considering almost-zero concentrations, indeed, if $k \to \infty$ then $c_0 = c_{lim}\exp(-k) \to 0$. However, the discussion driven in Ref. [3] (see Appendix A, Eq. (A.2)) now appears questionable. Indeed, the transition from the ionic-lattice concentration range down to the infinite-dilution limit should occur within an infinitesimal concentration range [0, $c_0$], where too few ions are present such that only almost-randomised lattices can actually be considered [5]. Perhaps, a better (and more convincing) argumentation about the case $k\to\infty$ is the following: the absence of point-charge carriers at any $c$ ($c_0 \leq c \leq c_{\lim}$) means that we are dealing with a (hypothetical) electrolyte where short-range interactions are so strong to determine the largest size of carriers at any (non-infinitesimal) concentration, such that $U$ becomes infinite because of the "infinite-size" of carriers (of course, the case $k\to\infty$ is only a mathematical abstraction). A consequence is that no long-range (ion-ion) interactions are detectable in the limit case $k\to\infty$, so their contribution to the total l.s.d. is null, which means $U_{I-I} \to 0$ at any (non-infinitesimal) concentration or, equivalently, from inspecting Eq. (24), $\alpha\to\infty$ when $k\to\infty$. Finally, collecting all limit conditions for the cases $k\to 1$ and $k\to\infty$, one has:

$$\alpha(k) \to \infty, k\to 1; \quad \frac{\sqrt{\alpha(k)}}{k} \to 0, k\to\infty; \quad \alpha(k) \to \infty, k\to\infty \tag{25}$$

Looking for $\alpha(k)$ such that all limits above are satisfied, it was analytically found [3] that

$$\sqrt{\frac{\pi\alpha(k)}{2}} = \frac{1}{C}\frac{k}{\ln(k)} \tag{26}$$

In Eq. (26) $C$ is a constant to determine. To the aim, it was assumed [3] that $U_{I-I}$ never exceeds $U_D$, such that $C=e^1$ (and, finally, Eq. (24)) was obtained after imposing that $k=e^1$ maximizes Eq. (26) when $U_{I-I}=U_D$. The assumption was based on the idea that $U_{I-I}=U_D$ should represent the case of the weakest ion-ion interactions, occurring when solute concentrations are almost zero. However, the idea was revised in Ref. [5], where Eq. (16) was derived, showing that the correct infinite-dilution expression for is $U_{I-I}=\frac{U_D}{\sqrt{\pi}}$.

Although so far proven consistent with real ionic lattices (we have always found $U_{I-I}<U_D$), there is need for further investigation about the assumption $U_{I-I}\leq U_D$, looking for a suitable ionic lattice corresponding to the case $k=e^1$, such that $\alpha(k=e^1)=2/\pi$, and $U_{I-I}=U_D$ at $k=e^1$.

In summary, combining Eqs. (19, 20, 24) allows writing:

$$U=\frac{U_D}{\sqrt{\pi}}\left[1+\frac{e^k c}{2k^2 c_{\lim}}\left(\frac{\pi e^1 \ln(k)}{k}(6k^2+1)-6k^2\right)-\frac{e^{2k}c^2}{k^2 c_{\lim}^2}\left(\frac{\pi e^1 \ln(k)}{k}(3k^2+1)-3k^2\right)+\frac{e^{3k}c^3}{2k^2 c_{\lim}^3}\left(\frac{\pi e^1 \ln(k)}{k}(2k^2+1)-2k^2\right)\right]$$

$$0\leq c \leq c_{\lim}\exp(-k) \tag{27a}$$

$$U=\frac{U_D}{\log_k\left(\frac{c_{\lim}}{c}\right)\left(\frac{c}{c_{\lim}}\right)^{1/k}}, \quad c_{\lim}\exp(-k)\leq c \leq c_{\lim} \tag{27b}$$

As already said, $k$ is obtained by solving Eq. (24), from the knowledge of the Madelung Constant $\alpha$ of the reference ionic lattice. For a given electrolyte, after setting the reference ionic lattice according to the electrolyte order, equation (27) provides an expression for $U$ at any concentration between 0 and $c_{lim}$ and it is fully determined provided that $c_{lim}$ is known (either experimentally or parametrically). Below some important aspects are discussed concerning the role of $c_{lim}$ and the meaning of the parameterisation based on this concentration.

## MEANING OF THE $c_{lim}$ PARAMETER IN THE QRL APPROACH

### Definition of $c_{lim}$

From Eq. (27b), if $c \to c_{lim}$ then $U \to +\infty$. Then, using Eq. (8) to calculate $\gamma_\pm$, if $c \to c_{lim}$ then $\ln(\gamma_\pm) \to 0$:

$$\ln(\gamma_\pm) \approx \frac{1}{k_B T} \frac{1}{2\nu N} \frac{1}{4\pi\varepsilon} \sum_{\substack{A,B \\ A \neq B}} \frac{Q_A Q_B}{R_{AB}} \frac{1}{\sqrt{\pi}} \frac{R_{AB}}{U} = \frac{1}{k_B T} \frac{1}{2\nu N} \frac{1}{4\sqrt{\pi}^3 \varepsilon} \frac{1}{U} (-\sum_A Q_A^2) \to 0 \quad (c \to c_{lim}) \quad (28)$$

In Eq. (28), the *erf* function is approximated with a usual first-order expansion [36], that is, $erf\left(\frac{R_{AB}}{2U}\right) \approx \frac{R_{AB}}{\sqrt{\pi} U}$ ($U \to +\infty$). Eq. (28) justifies the definition of $c_{lim}$ as the concentration corresponding to $\gamma_\pm = 1$ [1]. In the introductory remarks it was outlined that such a definition of $c_{lim}$ is consistent with measurements for a variety of systems, which is a very encouraging outcome of the QRL approach.

### Relationship with the Gibbs Free Energy

The QRL parameterisation shows applicative potentialities that can be enhanced by theoretical features of the model, so far evaluated only marginally. The main feature here considered is the connection of $c_{lim}$ with the Excess Gibbs Energy [46, 17, 18, 47]. First, we recall the relationship between $\gamma_\pm$ and the osmotic coefficient $\phi$, through the following form of the Gibbs-Duhem Equation [6]:

$$\Phi = 1 + \ln \gamma_\pm - \frac{1}{m} \int_0^m \ln \gamma_\pm \, dm \tag{29}$$

From Eq. (29) we also write

$$\int_0^m \ln \gamma_\pm \, dm = m(1 - \Phi + \ln \gamma_\pm) \tag{30}$$

Then, we shortly summarize main formulae expressing the Gibbs (Free) Energy for a binary solution [46, 17, 18, 47]. For convenience, hereinafter the symbol $n_{Sv}$ stands for number of solvent moles, and $n_s$ stands for number of solute moles. According to the Pitzer formalism [17, 18], the total Gibbs energy, $G^{id}$, of an ideal solution is defined as follows.

$$G^{id} = n_{Sv}(\mu^0_{Sv} - R_g T W_{Sv} \Sigma_i m_i) + n_s \Sigma_i \nu_i (\mu^0_i + R_g T \ln(m_i)) \tag{31}$$

In Eq. (31), $R_g$ is the Gas Constant, and $W_{Sv}$ indicates the molecular weight of solvent. For a binary electrolyte, either $v_i=v_+$ or $v_i=v_-$, and either $m_i=v_+m$ or $m_i=v_-m$. Concerning standard states and chemical potentials $\mu^0_{Sv}$ and $\mu^0_i$, $\mu^0_{Sv}$ refers to the pure solvent (at same $P$ and $T$, [46]), while $\mu^0_i$ are so chosen that the mean molal activity coefficient→1 when $m→0$ (at any $T$ and $P$ [46]). The total Gibbs energy, $G$, of a real solution expressed with molal activity coefficients of solute ions (below single-ion molal activity coefficients are indicated with $\gamma_i$, either $\gamma_i=\gamma_+$ or $\gamma_i=\gamma_-$) is given by [18, 47]:

$$G = n_{Sv}(\mu^0_{Sv} - R_g T W_{Sv}\Sigma_i m_i \Phi) + n_s \Sigma_i v_i [\mu^0_i + R_g T \ln(m_i \gamma_i)] \tag{32}$$

$G-G^{id}$ is the resulting total excess Gibbs energy, which can also be expressed in terms of $\ln(\gamma_\pm)$ and $\Phi$.

$$\frac{G - G^{id}}{R_G T} = W_{Sv} n_{Sv}\Sigma_i v_i m(1-\Phi) + n_s\Sigma_i v_i \ln(\gamma_i) = v n_{Sv} W_{Sv} m [(1-\Phi) + \ln(\gamma_\pm)] \tag{33}$$

Or, alternatively:

$$\frac{G - G^{id}}{R_G T} = n_s\Sigma_i v_i[(1-\Phi) + \ln(\gamma_i)] = v n_s [(1-\Phi) + \ln(\gamma_\pm)] \tag{34}$$

Throughout Eqs.(32-34), the relationships $m=n_s/(n_{Sv}W_{Sv})$ and $v_i m = m_i = \dfrac{n_i}{n_{Sv}W_{Sv}}$ (either $n_i=v_+ n_s$ or $n_i=v_- n_s$) has been used. Furthermore, partial derivatives below apply [47]:

$$\frac{\partial}{\partial n_{Sv}}\left(\frac{G-G^{id}}{R_G T}\right) = W_{Sv} v m (1-\Phi); \quad \frac{\partial}{\partial n_s}\left(\frac{G-G^{id}}{R_G T}\right) = v \ln\gamma_\pm \tag{35}$$

Equations above are obtained after considering the set of independent variables ($P$, $T$, $n_{Sv}$, $n_s$). Combining Eq. (30) with Eq.(33) also yields:

$$\int_0^m \ln\gamma_\pm\, dm = m(1-\Phi+\ln\gamma_\pm) = \frac{G-G^{id}}{v R_G T n_{Sv} W_{Sv}} \tag{36}$$

After considering the set of independent variables ($P$, $T$, $n_{Sv}$, $m$), while $n_s=m n_{Sv} W_{Sv}$, one obtains:

$$\frac{1}{v R_G T W_{Sv}}\frac{\partial}{\partial n_{Sv}}(G-G^{id}) = m(1-\Phi) \tag{37}$$

$$\frac{1}{v R_G T n_{Sv} W_{Sv}}\frac{\partial}{\partial m}(G-G^{id}) = \ln\gamma_\pm \tag{38}$$

With QRL, one has $c \leq c_{lim}$ and $\ln(\gamma_\pm) \leq 0$, with $\ln(\gamma_\pm)=0$ at $c=c_{lim}$[10]. From Eq. (38), the partial derivative of $G-G^{id}$ with respect to $m$ is negative for $0<c<c_{lim}$, and positive for $c>c_{lim}$. This allows concluding that $c=c_{lim}$ corresponds to a minimum[11] in the Excess Gibbs Energy (in the Pitzer form [17, 18]) with respect to $m$, provided that the other independent variables ($P$, $T$ and number of solvent moles $n_{Sv}$) are kept fixed.

---

[10] Other than at $c=0$.

Alternatively, after considering the set of independent variables ($P$, $T$, $n_{Sv}$, $n_s$), Eq. (35) states that $c=c_{lim}$ minimizes the Excess Gibbs Energy with respect to the number $n_s$ of solute moles, while keeping the other independent variables fixed.

*Relationship with Short-Range Interactions*

In Ref. [5], it was reconsidered the two-step QRL procedure [1] to deal with electrolyte solutions: firstly, the reference electrostatic configuration and secondly, the perturbing effects due to Brownian Motion in a thermal bath (first randomisation level in the model). Reconsideration yielded a generalization of the procedure so as to include the infinite dilution as a "limit" case where the reference configuration is no longer electrostatically ordered (ions are very few even at microscopic scale), but rather represented by an almost-randomised lattice (second randomisation level). In the infinite and strong dilution cases, the reference configuration corresponds to ions at the reticular points in their final state, to be intended as the most observed configuration after multiple repetitions of the same process. The starting state is chosen such that ions are initially motionless and infinitely apart from each other such that their coulombic interaction energy is null, and the initial configuration is represented by a fully random lattice. The analogy with the Debye charging process [6, 46] is evident. The total (kinetic+potential) energy of the $\nu$N-particle system is evidently null. Considering here the strong dilution, for any molecule[12] the energy condition becomes [5]:

$$\sum_{i=1}^{\nu} \frac{M_i v_{el,i}^2}{2} + \frac{\alpha Q_+ Q_-}{4\pi\varepsilon R} = 0 \qquad \text{(Strong Dilution)} \qquad (39)$$

In Eq. (39), $M_i$ and $\mathbf{v}_{el,i}$ are the mass and velocity of ion $i$ in presence of electrostatic forces only (for brevity, the symbol "$i$" will be hereinafter omitted). Equation above is based upon the assumption that short-range interactions among ions are negligible until $R$ is sufficiently large, and the solvent can be considered as a "continuous medium". Since $R$ is large and velocities are (in modulus) of the order of $1/R^{1/2}$, we can speak about "slow-dynamic changes", and almost-zero velocities that allow for keeping a reference configuration for a long while (in spite of the minimum of potential energy at $R=0$).

According to the second step of the QRL procedure, thermal effects are then included in the form of thermal noise and dissipative collisions among ions and molecules, i.e., in terms of Brownian Motion and viscous forces. The probability $\mathscr{P}$ for an ion to remain within its own lattice cell is definitely reduced by such thermal effects, and estimated [1] by means of the rate between electrostatic and thermal velocities, which yielded the following relationship[13] with the l.s.d. $U$ at strong dilution [1, 5]:

---

[11] In the present context, for brevity it is supposed that saturation will occur at $c_{sat} > c_{lim}$.
[12] According to Ref. [5], the term molecule hereinafter indicates a generic grouping of $\nu_+$ positive and $\nu_-$ negative ions.
[13] The relationship is strictly valid at $c=c_{lim}\exp(-k)$, however, for this paragraph, it can also be used in the general case of strong dilution.

$$\mathscr{P}^{1/3} = \frac{v_{el}}{<v_{th}>} = \frac{\sqrt{\frac{2\alpha|Q_+Q_-|}{M4\pi\varepsilon R}}}{\sqrt{\frac{2k_BT}{\pi M}}} = \sqrt{\frac{\alpha|Q_+Q_-|}{4\varepsilon Rk_BT}} = \sqrt{\frac{\alpha|Q_+Q_-|\rho vR^2}{4\varepsilon k_BT}} = \frac{R}{\sqrt{2\pi}\left(\frac{\sqrt{2}U_D}{\sqrt{\pi\alpha}}\right)} = \frac{R}{\sqrt{2\pi}U} \quad (40)$$

In Eq. (40) $v_{el}$ indicates the velocity (in modulus) pertaining a given ion in a lattice cell[14]. When solutions are not strongly dilute, neglecting short-range interactions is evidently wrong. However, the process for which ions are displaced from their initial free state to a final lattice can still be performed provided that Eq. (39) is corrected so as to include short-range forces and potential, say $V_{S-R}$, such that:

$$\sum_{i=1}^{\nu}\frac{M_iv_{el,i}^2}{2} + \frac{\alpha Q_+Q_-}{4\pi\varepsilon R} + V_{S-R} = 0 \quad (41)$$

In analogy with Eq. (40), we may then write:

$$\mathscr{P}^{1/3} = \frac{v_{el}}{<v_{th}>} = \frac{\sqrt{\frac{2}{M}\left(\frac{\alpha|Q_+Q_-|}{4\pi\varepsilon R} - V_{S-R}\right)}}{\sqrt{\frac{2k_BT}{\pi M}}} = \frac{\sqrt{\left(\frac{\alpha|Q_+Q_-|}{4\varepsilon R} - \pi V_{S-R}\right)}}{\sqrt{k_BT}} = \sqrt{\left(\frac{\alpha|Q_+Q_-|}{4\varepsilon Rk_BT} - \frac{\pi V_{S-R}}{k_BT}\right)} =$$

$$= \sqrt{\frac{\alpha|Q_+Q_-|}{4\varepsilon Rk_BT}\left(1 - \frac{4\pi\varepsilon RV_{S-R}}{\alpha|Q_+Q_-|}\right)} = \frac{R}{\sqrt{2\pi}\left(\frac{\sqrt{2}U_D}{\sqrt{\pi\alpha}}\right)}\sqrt{\left(1 - \frac{4\pi\varepsilon RV_{S-R}}{\alpha|Q_+Q_-|}\right)} = \frac{R}{\sqrt{2\pi}U} \quad (42)$$

Eq. (42) may still represent the probability $\mathscr{P}^{1/3}$ resulting from the spatial integration of a Gaussian Density (along with a given spatial axis, from $-R/2$ to $R/2$) provided that $R$ is still small compared with $U$. The condition $R<<U$ is exactly satisfied in the limit $c\rightarrow c_{lim}$. From comparison between Eq. (42) and Eq. (40) we immediately see that

$$\sqrt{\left(1 - \frac{4\pi\varepsilon RV_{S-R}}{\alpha|Q_+Q_-|}\right)} = \exp(1)\frac{1}{k}\ln\left(\frac{c_{lim}}{c}\right)\left(\frac{c}{c_{lim}}\right)^{\frac{1}{k}} = \exp(1)\frac{3}{k}\ln\left(\frac{R}{R_{lim}}\right)\left(\frac{R}{R_{lim}}\right)^{\frac{3}{k}} \quad (c\rightarrow c_{lim}) \quad (43)$$

In the last member of Eq. (43), the mean distance $R_{lim}$ (corresponding to $c_{lim}$) has been evidenced (if $c\leq c_{lim}$ then $R\geq R_{lim}$), and usual relationships between $c$, N and $R$ (Eq. (1)) were used. Then:

$$V_{S-R} = \frac{\alpha|Q_+Q_-|}{4\pi\varepsilon R}\left(1 - \left(\exp(1)\frac{3}{k}\ln\left(\frac{R}{R_{lim}}\right)\left(\frac{R}{R_{lim}}\right)^{\frac{3}{k}}\right)^2\right) \quad (c\rightarrow c_{lim}) \quad (44)$$

---

[14] In a maximizing condition for one velocity component while the others are zero, from Eq. (39), and supposing that all remaining ions in the "molecule" are kept motionless.

**CONCLUDING REMARKS: FURTHER RESEARCH UNDER DEVELOPMENT**

*Further Generalization of the Model*

QRL includes reference lattices to deal with electrolyte solutions. For any given electrolyte-order, the lattice is set so as to describe the solution volume with identical cells. Introduction of volumetric distortion (that is, lattice cells with different volumes), might be useful to account for high-order asymmetry effects that are difficult to represent only using regular space-group geometries. In this connection, it might be helpful introducing geometrical distortion in the reference lattice even in the case of ion-association at strong dilution. In the past [2], di-divalent metal-sulphates were investigated, and a procedure was developed based on the inclusion of association constants, in order to account for ion-association at strong dilution. The advantage offered by a geometry-based approach would be advancing towards a unified pseudo-lattice theory, available for all cases.

*Calculation of Volumetric and Thermal Properties with QRL*

QRL is based on the modelling of $\ln(\gamma_\pm)$ so, the most natural route for calculating (partial and apparent) molar volumes, enthalpies and capacities passes through pertinent partial derivatives of $\ln(\gamma_\pm)$ with respect to $P$ and $T$ [46]:

$$\left.\frac{\partial \ln \gamma_\pm}{\partial P}\right|_{m,T} = \frac{\overline{V}_2 - \overline{V}_2^0}{\nu R_g T}; \qquad \left.\frac{\partial \ln \gamma_\pm}{\partial T}\right|_{m,P} = -\frac{\overline{L}_2}{\nu R_g T^2}; \qquad -\left.\frac{\partial}{\partial T}\left(\nu R_g T^2 \frac{\partial \ln \gamma_\pm}{\partial T}\right)\right|_{m,P} = \overline{C}_2 - \overline{C}_2^0 \quad (45)$$

In Eq. (45), partial molal volume of solute[15] $\overline{V}_2$ and corresponding Standard State value (labelled with the apex 0) appear. Similarly for partial molal enthalpy $\overline{L}_2$ and capacity $\overline{C}_2$, and their corresponding Standard State values[16]. Apparent properties can then be obtained by means of the appropriate integral equations after calculating the pertinent partial properties [46]. Using QRL, partial derivatives can in principle be calculated starting from Eq. (8), (the procedure is almost straightforward, however rather long and a bit tedious), yielding appropriate expressions (not here reported for brevity) for each thermodynamic property, provided that Standard State values are known. In this context, however, the main focus is on the fact that such derivatives will involve $c_{\lim}$ as well as its derivatives with respect to $P$ or $T$. So, calculation of $\overline{V}_2$ will involve $c_{\lim}$ and $\frac{\partial c_{\lim}}{\partial P}$; calculation of $\overline{L}_2$ will involve $c_{\lim}$ and $\frac{\partial c_{\lim}}{\partial T}$; calculation of $\overline{C}_2$ will involve $c_{\lim}$, $\dfrac{\partial c_{\lim}}{\partial T}$ and $\dfrac{\partial^2 c_{\lim}}{\partial T^2}$. A way to obtain partial derivatives of $c_{\lim}$ passes through the partial derivatives of $m_{\lim}$, in the $mPT$ frame, and through the fact that, since $\ln(\gamma_\pm(m_{\lim}, P, T)) = 0$ for any $P$ and $T$, any total derivative (with respect to either $P$ or $T$) is zero at $m = m_{\lim}(P, T)$.

---

[15] The solute is usually indicated with 2, and the solvent with 1.
[16] Partial derivatives with respect to solute or solvent densities (and their relationships with Kirkwood-Buff Integrals and isothermal compressibility [41, 42, 48]), though not here explicitly recalled, might also be considered.

$$\left.\frac{d(\ln\gamma_\pm)}{dP}\right|_{m=m_{\lim}} = 0 \Leftrightarrow \left.\frac{\partial(\ln\gamma_\pm)}{\partial P}\right|_{m=m_{\lim}} + \frac{\partial m_{\lim}}{\partial P}\left.\frac{\partial(\ln\gamma_\pm)}{\partial m}\right|_{m=m_{\lim}} = 0 \qquad (46a)$$

$$\left.\frac{d(\ln\gamma_\pm)}{dT}\right|_{m=m_{\lim}} = 0 \Leftrightarrow \left.\frac{\partial(\ln\gamma_\pm)}{\partial T}\right|_{m=m_{\lim}} + \frac{\partial m_{\lim}}{\partial T}\left.\frac{\partial(\ln\gamma_\pm)}{\partial m}\right|_{m=m_{\lim}} = 0 \qquad (46b)$$

$$\left.\frac{d^2(\ln\gamma_\pm)}{dT^2}\right|_{m=m_{\lim}} = 0 \Leftrightarrow \left.\frac{\partial^2(\ln\gamma_\pm)}{\partial T^2}\right|_{m=m_{\lim}} + 2\frac{\partial m_{\lim}}{\partial T}\left.\frac{\partial^2(\ln\gamma_\pm)}{\partial m\partial T}\right|_{m=m_{\lim}} + \left(\frac{\partial m_{\lim}}{\partial T}\right)^2\left.\frac{\partial^2(\ln\gamma_\pm)}{\partial m^2}\right|_{m=m_{\lim}} + \frac{\partial^2 m_{\lim}}{\partial T^2}\frac{\partial(\ln\gamma_\pm)}{\partial m} = 0 \qquad (46c)$$

Concerning, e.g., first-order derivatives[17], manipulating Eqs. (45-46) gives:

$$\frac{\partial m_{\lim}}{\partial P} = -\frac{\dfrac{\overline{V}_2\big|_{m=m_{\lim}} - \overline{V}_2^{\,0}}{\nu R_g T}}{\left.\dfrac{\partial \ln(\gamma_\pm)}{\partial m}\right|_{m=m_{\lim}}} \quad ; \quad \frac{\partial m_{\lim}}{\partial T} = \frac{\dfrac{\overline{L}_2\big|_{m=m_{\lim}}}{\nu R_g T^2}}{\left.\dfrac{\partial \ln(\gamma_\pm)}{\partial m}\right|_{m=m_{\lim}}} \qquad (47)$$

So, at least in principle, wished derivatives can be calculated from the knowledge of partial properties at $m=m_{\lim}$, at any given $P$ and $T$. Note also that QRL needs for converting from molal to molar scale at any $P$ and $T$, which means to know derivatives of solution densities with respect to $T$ and $P$ (although simulations so far performed are showing that such derivatives, if experimentally unavailable, can be replaced by those of pure solvent, with a very little influence on final results). In addition, experimental information about volumetric and thermal partial properties, needed to determine the QRL parameters, must be accurate. Partial properties are known to be difficult to measure, so it might better be exploring relationships of $m_{\lim}$ and its derivatives with apparent properties. Although preliminarily, implementation of Eq. (45) by means of partial derivatives of Eq. (8) has so far pointed out that results qualitatively respect the experimental trends of investigated cases. Their accuracy is, however, quantitatively inferior to that currently available from many-parameter models (e.g., from Pitzer Theory). In particular, this is true for thermal properties. Hopefully, current simulations are suggesting that accuracy can definitely be increased up to a competitive level by introducing very few additional parameters (other than $c_{\lim}$ and its derivatives). An advantage of these parameters is that they are again related with concentrations, without exceeding in complicating the model.

In summary, additional parameterisation (although very limited compared to common many-parameter models [8, 15]) and complexity of experimental procedures for determining partial properties are, at present, limitations for a practical application of QRL. Alternative or mixed strategies should also be explored. A very appealing approach, currently under investigation, is based on the fact that, after all, derivatives of $c_{\lim}$ with respect to $T$ or $P$ might be obtained from the knowledge of $c_{\lim}$ values at $T\pm\Delta T$, $T\pm 2\Delta T$, $P\pm\Delta P$, $\Delta T$ and $\Delta P$ suitably chosen (for example, $\Delta T = 5°C$; $\Delta P = 1\div 5$ MPa). Determination of $c_{\lim}$ (or $m_{\lim}$) values at different pressures and temperatures can be made by means of measurements of $\ln(\gamma_\pm)$ at such pressures and temperatures, looking for the concentration such that $\ln(\gamma_\pm)=0$. It is possible strongly reducing the number of needed measurements by observing that, according to QRL, $\ln(\gamma_\pm)$ depends on $c/c_{\lim}$ at any $c$. So, an iterative procedure might be as follows: if $c=c^*$ is the concentration used at a given search-step, then an estimate-step of $c_{\lim}$, say $c_{\lim}^*$, can be done by imposing $c_{\lim}^*$ such that measured $\ln(\gamma_\pm)$ (when negative) will be equal to that calculated at $c=c^*$. This procedure normally converges quickly to $c_{\lim}^*=c_{\lim}$, also considering that differences between measured and calculated $\ln(\gamma_\pm)$ are vanishing if $c\rightarrow c_{\lim}$.

---

[17] We here omit explicit expression for second-order derivatives of $m_{\lim}$ with respect $T$, for brevity.

*Generalization to Multi-Solvent or Multi-Salt Solutions*

A way to deal with multi-solvent solutions is suggested by the Mixture Theory [49, 27, 50], which includes various equations to model the dielectric permittivities of mixed solvents at a mesoscopic scale. A way to deal with multi-salt solutions is suggested by the observation that contributions to Eq (8) from the effective-carrier population can be shared according to the nominal charge of the effective-carriers, such that it might also be possible, at least in principle, to explore single-ion activities and related equations for multi-salt solutions [46]. These points are currently under investigation.

*Thermodynamic Consistency of QRL*

In Ref. [5] it was shown that the Zero-Order Moment Equation [7] is automatically satisfied by QRL, provided that effective-carriers are taken into account. However, while carriers correspond to ions and effective –carriers are relatively easy to interpret at strong-dilution [1, 5], their nature (and interpretation) is less immediate at higher concentrations. This research-point is still under investigation, as well as the evaluation of consistency of the model with respect to higher-order Moment Equations [7]. More generally, concerning very popular debates upon thermodynamic routes and consistency of models with respect to them [6-8], it is here to outline that with QRL consistency is better evaluated through comparison with experimental results, since the model was developed within the LR frame. Kirkwood-Buff Integrals [41, 42, 48] are also tractable with QRL.